\documentclass{iau}

\usepackage{amsmath}
\usepackage{graphicx}
\usepackage{multirow}

\begin{document}

\lefttitle{E. da Cunha}
\righttitle{The dust properties of star-forming galaxies in the first billion years}

\jnlPage{1}{7}
\jnlDoiYr{2022}
\doival{10.1017/xxxxx}
\volno{373}

\aopheadtitle{Proceedings IAU Symposium}

\title{The dust properties of star-forming galaxies in the first billion years}

\author{Elisabete da Cunha}
\affiliation{International Centre for Radio Astronomy Research, University of Western Australia, 35 Stirling Hwy, Crawley, WA 6009, Australia}
\affiliation{ARC Centre of Excellence for All Sky Astrophysics in 3 Dimensions (ASTRO 3D)}

\begin{abstract}
The Atacama Large Millimetre/Sub-millimetre Array (ALMA) is obtaining the deepest observations of early galaxies ever achieved at (sub-)millimetre wavelengths, and detecting the dust emission of young galaxies in the first billion years of cosmic history, well in the epoch of reionization. Here I review some of the latest results from these observations, with special focus on the REBELS large programme, which targets a sample of 40 star-forming galaxies at $z\simeq7$. ALMA detects significant amounts of dust in very young galaxies, and this dust might have different properties to dust in lower-redshift galaxies. I describe the evidence for this, and discuss theoretical/modelling efforts to explain the dust properties of these young galaxies. Finally, I describe two additional surprising results to come out of the REBELS survey: (i) a new population of completely dust-obscured galaxies at $z\simeq7$, and (ii) the prevalence of spatial offsets between the ultraviolet and infrared emission of UV-bright, high-redshift star-forming galaxies.
\end{abstract}


\maketitle

\section{Introduction}

Interstellar dust plays a critical role in galaxy assembly. Dust grains are a site of efficient molecule formation, specifically H$_2$, the fuel for new star formation (e.g., \citealt{Hirashita2002}). Dust grains also affect how we observe galaxy assembly: the cosmic infrared background is equivalent in power to the cosmic optical background (e.g., \citealt{Dole2006}), implying that about half of all the UV/optical light produced by stars in the Universe is reprocessed by dust. In fact, we know that dust-obscured star formation contributes significantly to the total star formation rate (SFR) density of the Universe (e.g., \citealt{Madau2014}). Recent studies show that dust-obscured star formation dominates the cosmic SFR density out to at least $z\simeq3$ (e.g., \citealt{Zavala2021}), and it still provides an important contribution at higher redshifts, although the exact amount of dust-obscured star formation at the earliest redshifts ($z\gtrsim6$, i.e., in the epoch of reionization, or `cosmic dawn') is still uncertain.

The main limitation in measuring obscured star formation at the highest redshifts has been the difficulty to detect and measure the dust emission of individual high-redshift galaxies using far-infrared/sub-millimetre facilities such as the {\em Herschel Space Observatory} and single-dish sub-mm observatories. Due to their poor resolution and sensitivity (see, e.g., figure 10 of \citealt{Casey2014}), observations with these facilities suffer from important confusion noise, which limited the study of high-redshift sources to only the brightest objects, such as quasars (e.g., \citealt{Bertoldi2003}) and sub-millimetre galaxies (SMGs; e.g., \citealt{Walter2012,Riechers2013}). Alternatively, several studies attempted to overcome the confusion limits and detect fainter galaxies using stacking (e.g., \citealt{Schreiber2015}) and de-blending techniques (e.g., \citealt{Liu2018,Jin2018}), but these approaches have limitations. With its unprecedented sensitivity and spatial resolution at (sub-)millimetre wavelengths, the Atacama Large Millimetre/Sub-millimetre Array (ALMA), has revolutionised this field, allowing us to directly detect, for the first time, the dust (and gas) emission from individual, normal star-forming galaxies at $z>6$ (see \citealt{Hodge2020} for a recent review). In this paper, I briefly review what recent ALMA observations are revealing about the interstellar dust properties of star-forming galaxies in the first billion years.

\section{ALMA observations of Lyman-break galaxies}

\subsection{Different dust properties at high-redshift?}

\begin{figure*}[t]
\begin{center}
\includegraphics[width=\textwidth]{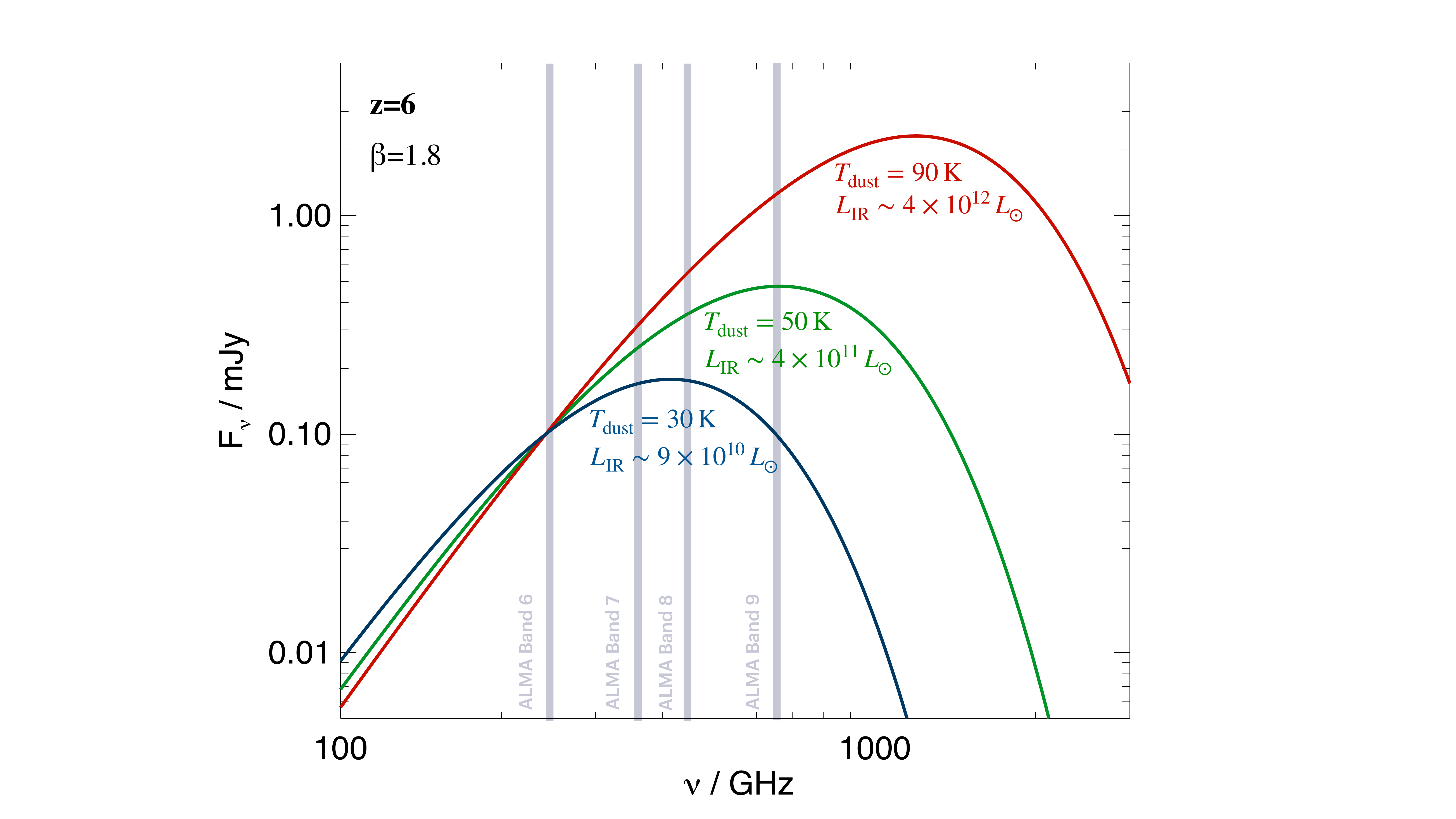}
\caption{Three model dust emission spectral energy distributions (SEDs) at $z=6$ normalized at the same ALMA Band 6 (230 GHz) flux of 0.1 mJy. This illustrates how sources with the same Band 6 flux can have inferred infrared luminosities, $L_\mathrm{IR}$, that vary by almost two orders of magnitude depending on the adopted dust temperature, $T_\mathrm{dust}$. To constrain the temperatures and luminosities, we need multi-band observations at higher frequencies (e.g., in the ALMA bands shown in grey) sampling the dust SED closer to its peak.}
\label{sed}
\end{center}
\end{figure*}

Before direct dust detections with ALMA, high-redshift studies routinely corrected the observed UV emission of distant galaxies for dust attenuation using the so-called `IRX-$\beta$' relation between the ultraviolet spectral slope ($\beta$) and the infrared excess (IRX, defined as the ratio of infrared to ultraviolet luminosity). Local starburst galaxies follow a tight relation in the IRX vs $\beta$ diagram \citep{Meurer1999}, which can be explained if stars are behind a screen of dust with properties similar to those of dust in the Milky Way \citep{Calzetti1994}. Before the advent of ALMA, high-redshift galaxies were assumed to follow a similar relation. However, early ALMA follow-up of known Lyman-break galaxies (LBGs) at $z>4$ in Band 6 ($\simeq1$\,mm) found infrared luminosities that were lower than those predicted by the local starburst IRX-$\beta$ relation. This was found in a variety of targeted observations and stacks of $z>4$ LBGs (e.g., \citealt{Capak2015,Bouwens2016,Fudamoto2017,Barisic2017,Bowler2018,Schouws2022}): the lower IR excesses were more consistent with a steeper dust extinction curve similar, to the one observed in the Small Magellanic Cloud. Taken at face value, this result would imply that interstellar dust in EoR galaxies has fundamentally different properties than dust in low-redshift galaxies. This led to speculation that perhaps these galaxies have different dust properties due to their lower metallicities. This also implied that dust properties evolve rapidly in the early universe, since studies at $z<4$ seemed to show properties already consistent to Milky Way dust (e.g., \citealt{Fudamoto2020}). However, these results are plagued by the large uncertainty in measuring the total infrared luminosity of a galaxy from a single 1-mm ALMA measurement without knowing the dust temperature (see Fig.~\ref{sed} and discussion in section 3.3 of \citealt{Hodge2020}).

\subsection{The need for robust dust temperatures}

In all the aforementioned studies, a dust temperature had to be assumed to extrapolate the single-band ALMA measurement to a total IR luminosity, and most studies assumed temperatures around 40~K, similar to local and intermediate-redshift star-forming galaxies. \cite{Faisst2017} argued that local analogues to high-redshift LBGs have higher dust temperatures ($T_\mathrm{dust} \sim 60-90\,\mathrm{K}$), and that adopting such temperatures would bring the ALMA detections (and upper limits) more in agreement with Milky-Way dust in the IRX-$\beta$ relation. Theoretical work also seems to support hotter dust at high redshifts (e.g., \citealt{Behrens2018,Narayanan2018,Ma2019,Liang2019,McAlpine2019,Sommovigo2020}).

\begin{figure*}
\begin{center}
\includegraphics[width=\textwidth, trim={2cm 0cm 2cm 0cm}]{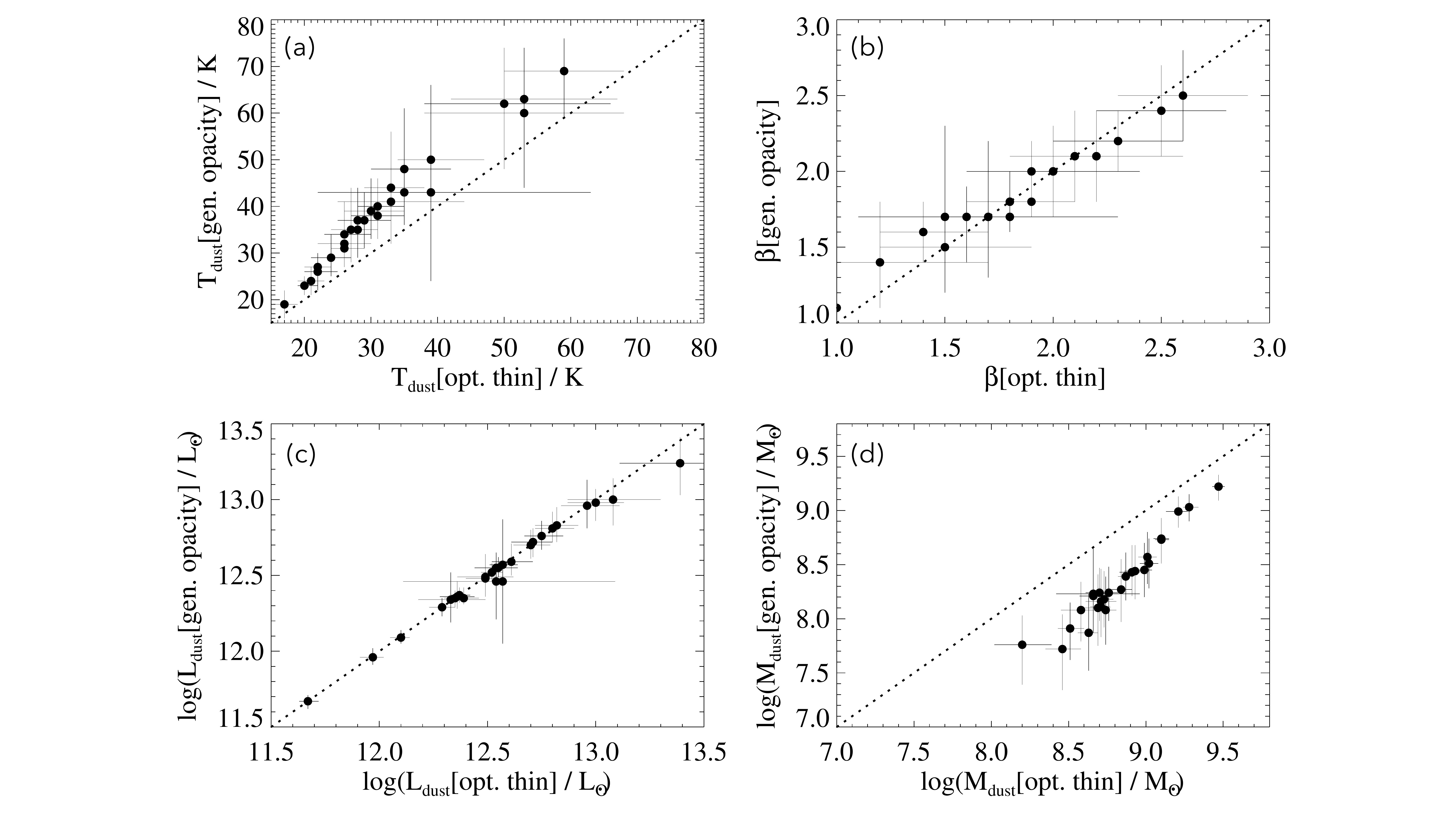}
\caption{Comparison between dust physical properties derived from fitting the dust emission with simple isothermal models when using an optically-thin assumption ($x$-axes) vs a general opacity model ($y$-axes): (a) dust temperature; (b) dust emissivity index; (c) dust luminosity; (d) dust mass. The points shown are 27 SMGs with well-sampled dust SEDs studied in \cite{daCunha2021}. While the dust luminosity and emissivity index are quite robust to optical depth assumptions, assuming the wrong optical depth can introduce biases in the derived dust temperatures and dust masses (see more details in \citealt{daCunha2021}). Figure first published in \cite{daCunha2021}. }
\label{sed_uncertainties}
\end{center}
\end{figure*}

These higher dust temperatures need to be confirmed through multi-frequency observations with ALMA, preferably sampling close to the peak of the dust emission, to measure $T_\mathrm{dust}$. For galaxies at $z\gtrsim6$, such observations are time consuming even with ALMA, but they have been achieved for a handful of galaxies so far. \cite{Faisst2020} followed-up four $z\sim5.5$ galaxies in Band 8 (sampling rest-frame $\sim100\mu$m, close to the peak of dust emission), which, combined with existing Bands 6 and 7 observations, allowed them to measure dust temperatures $T_\mathrm{dust} \sim 30-40\,\mathrm{K}$. In another recent study, \cite{Bakx2021} combined Band 9 observations with observations in three other bands (Bands 6, 7, and 8), to constrain a dust temperature of $T_\mathrm{dust}=42\,\mathrm{K}$ for a galaxy at $z=7.13$.

It is worth noting that having ALMA high-frequency observations that sample the peak of the dust emission (or even beyond the peak) is critical for robust constraints on the dust temperature (and hence, luminosity). But even with multi-band observations, potential degeneracies with the dust emissivity index and dust opacity assumptions need to be taken into account (e.g., \citealt{daCunha2021,Jin2022,Witstok2022}). For example, in a detailed study of the dust emission of SMGs, \cite{daCunha2021} show that with only three ALMA bands sampling the dust emission SED, it is very difficult to distinguish between optically thin dust emission and dust that it optically thick at far-infrared wavelengths -- but important systematics can be introduced in the derived dust temperatures and masses (Fig.~\ref{sed_uncertainties}).

\subsection{Is dust in galaxies really getting hotter with redshift?}

As mentioned before, theoretical simulations predict an increase of the dust temperature with redshift, which is generally explained by several factors such as stronger radiation fields, lower dust masses, and more compact dust-emitting regions at high redshifts. Empirically, an increase of dust temperature with redshift has also been observed, e.g., by \cite{Schreiber2018}, though we note that this is still somewhat controversial as it could be a result of selection effects: since there is a positive correlation between dust temperature and dust luminosity, and flux-limited samples detecting higher luminosity sources at higher redshifts (e.g., \citealt{Ugne2020}). In fact, the direct temperature measurements of sources at $z\gtrsim6$ with multi-band ALMA observations described above \citep{Faisst2020,Bakx2021} do not follow the extrapolation of the temperature-redshift relation of \cite{Schreiber2018}: their temperatures are about 10 to 20 K lower than predicted by that relation. If these temperatures are typical for high-redshift galaxies, then the dust temperatures might not be high enough to reconcile the IR excesses with Milky Way-like dust attenuation in the IRX-$\beta$ relation, which would indicate that indeed galaxies at the EoR have different dust properties. However, it is to be expected that there is a spread in dust temperatures at any given redshift (depending on different galaxy properties), so larger samples of galaxies t $z\gtrsim6$ with ALMA continuum observations are needed. One such samples comes from the REBELS survey described below.

\section{Results from the REBELS survey}

\subsection{Continuum detections in REBELS}

\begin{figure*}[t]
\begin{center}
\includegraphics[width=\textwidth,trim={0cm 1cm 0cm 0cm}]{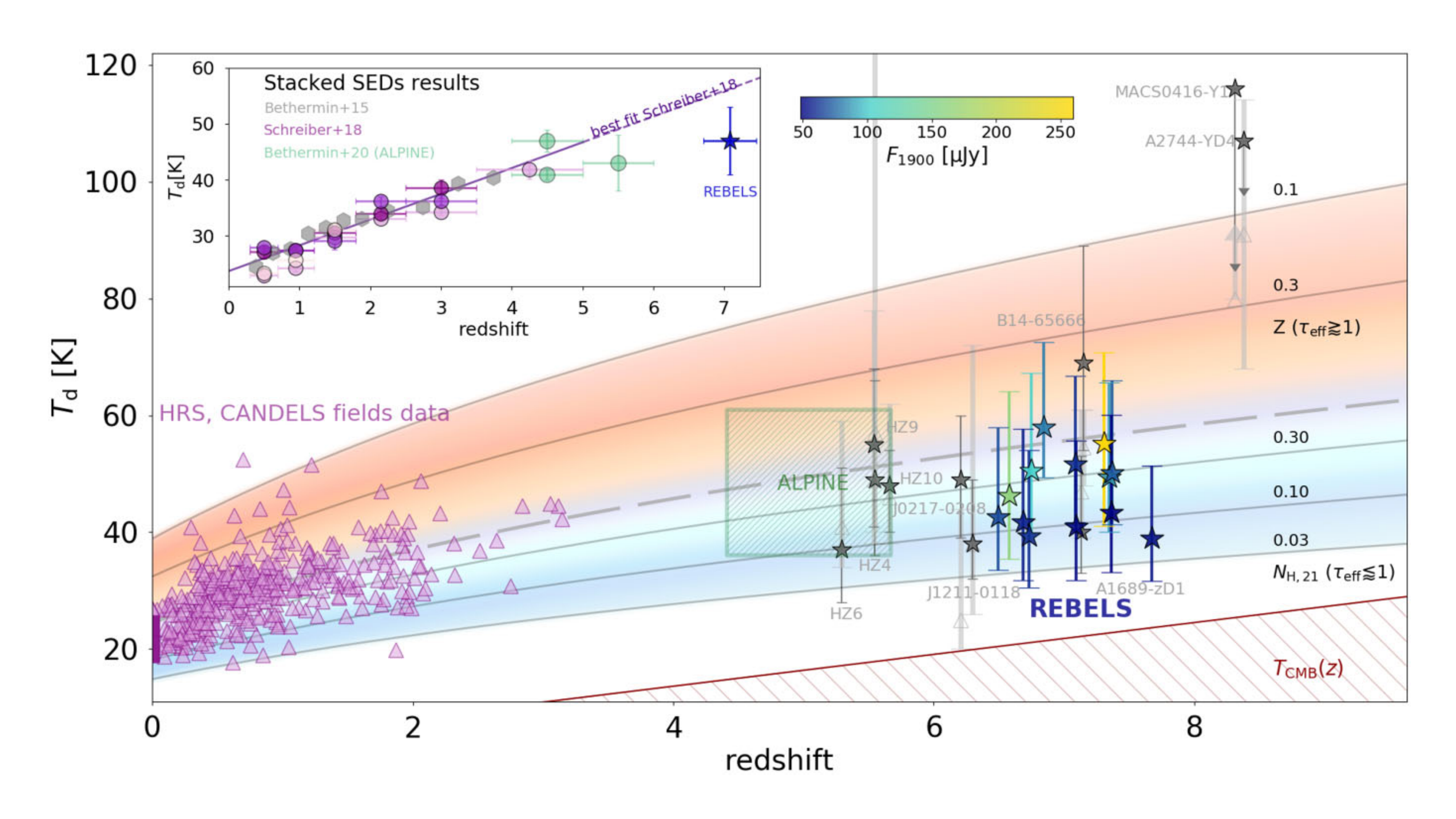}
\caption{Empirical relation between the redshift and the dust temperature of galaxies. The REBELS sources are plotted as colourful stars. The inset shows that the average dust temperature of REBELS sources fall short of the extrapolation of the $T_\mathrm{dust}$-redshift relation of \cite{Schreiber2018}. Figure first published in \cite{Sommovigo2022}.}
\label{tdustz}
\end{center}
\end{figure*}

The Reionization Epoch Bright Emission Line Survey (REBELS) is an ALMA Large Programme targeting [OIII] and [CII] lines and their underlying dust continuum (at $\sim90$ and $160\mu$m) in 40 UV-bright LBGs at $z>6.5$ \citep{Bouwens2022}. From this sample, 18 sources have been detected in the continuum so far \citep{Inami2022}, which more than tripled the previous number of dust continuum detections at $z>6.5$. The inferred infrared luminosities of these continuum-detected sources imply that about half of their star formation is dust-obscured. One of the detections, REBELS-25, is noteworthy as it is an ultra-luminous infrared galaxy (ULIRG, with $L_\mathrm{IR}>10^{12}\,L_\odot$) at $z=7.4$ (Hygate et al., submitted).

So far the REBELS sources are only continuum-detected in one ALMA band, therefore we cannot directly measure the dust temperature from fitting the dust SEDs. But \cite{Sommovigo2022} estimate the temperatures of REBELS sources using a novel analytical method that uses only the [CII] line and its underlying dust continuum \citep{Sommovigo2021}. They find an average $T_\mathrm{dust}\simeq46\,\mathrm{K}$ for the REBELS sources (Fig.~\ref{tdustz}) -- broadly consistent with the more direct multi-band measurements of \cite{Faisst2020,Bakx2021} for galaxies at similar redshifts.

\subsection{Dust-obscured star formation and its contribution to the cosmic SFR density at $z\simeq7$}

\begin{figure*}[t]
\begin{center}
\includegraphics[width=\textwidth, trim={1cm 3.5cm 1cm 2cm}]{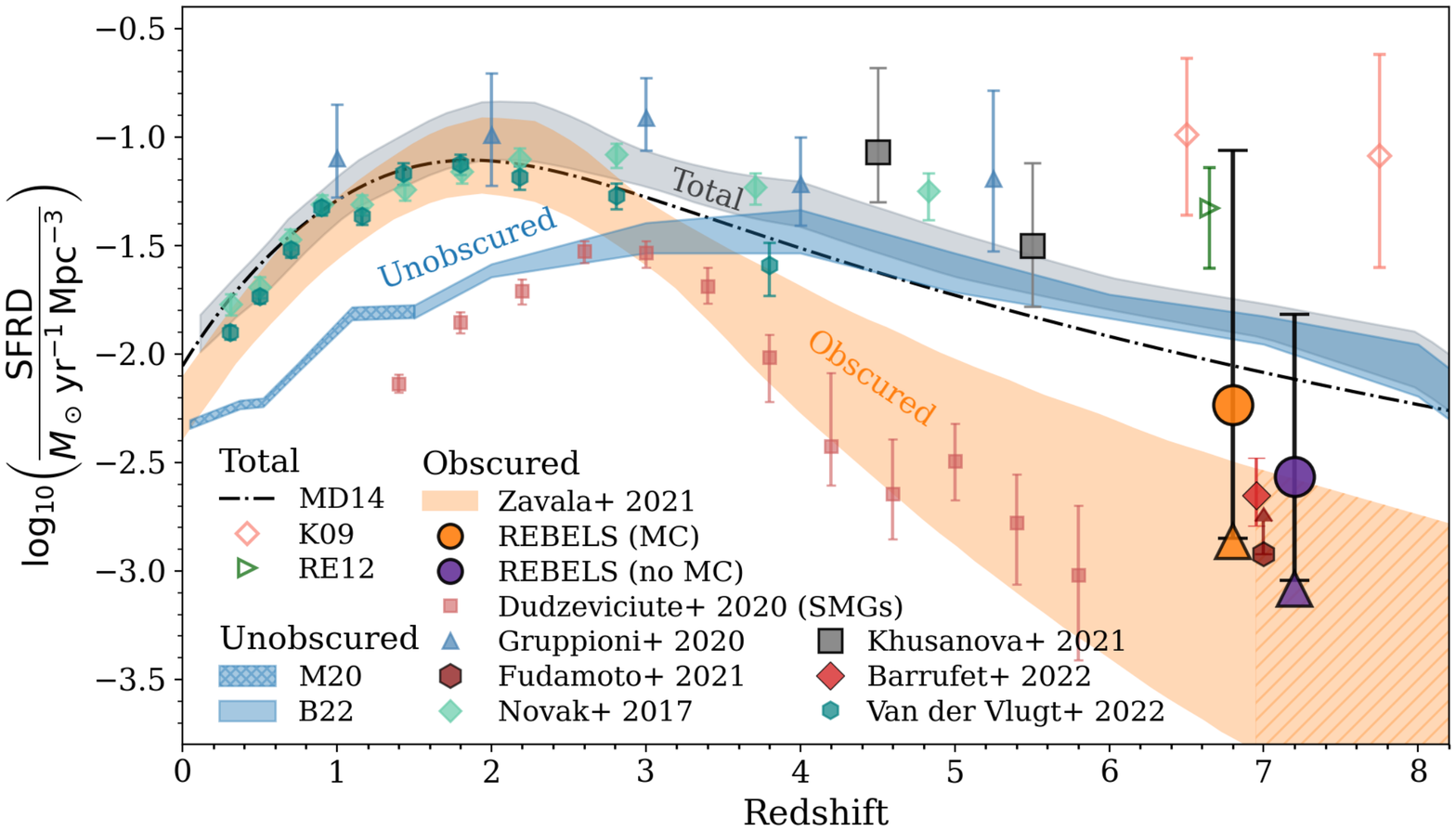}
\caption{Redshift evolution of the cosmic star formation rate density. The purple and orange filled circles/triangles show the estimated contribution by the REBELS sources at $z\simeq7$. Figure first published in \cite{Algera2022}.}
\label{csfrd}
\end{center}
\end{figure*}

The dust temperature estimates of REBELS sources by \cite{Sommovigo2022} are used to extrapolate from the single continuum measurements to obtain their total infrared luminosities, and infer their dust-obscured star formation rates. \cite{Algera2022} find that between 30\% and 60\% of the star formation rate in these galaxies is obscured. They find that this fraction depends on the stellar mass of the sources, similar to what is found at lower redshifts (\citealt{Whitaker2017}; the fraction of dust-obscured SFR increases with stellar mass). They also use these results to quantify the contribution of dust-obscured star formation to the cosmic star formation rate density at $z\sim7$. Since REBELS is not a blind, volume-limited survey, but a targeted survey, infrared luminosity functions cannot be directly directly derived. Their method involves first finding a relation between the dust luminosity and stellar mass, and then integrating across the (known) stellar mass function at $z=7$ (\citealt{Algera2022}; see also, Barrufet et al., submitted). \cite{Algera2022} find that dust-obscured star formation at high redshifts is more important than previously assumed: it can contribute up to a third of the total cosmic SFR density at $z\simeq7$ (Fig.~\ref{csfrd}). This implies also that the cosmic SFR density does not decrease as fast as previously predicted at $z>3$. While there are caveats attached to this result (stellar mass and dust luminosity uncertainties, sample selection effects, etc.), it could have a profound meaning: that galaxies in the epoch of reionization already have significant amounts of dust, and indeed perhaps more dust than we were expecting to observe in such young galaxies. This in turn might challenge our understanding of how metals and dust evolve in the interstellar medium of early galaxies, as I discuss in the next section.

\section{Can we explain so much dust at early times?}

\begin{figure*}[t]
\begin{center}
\includegraphics[width=\textwidth,trim={1cm 0cm 3.5cm 0cm}]{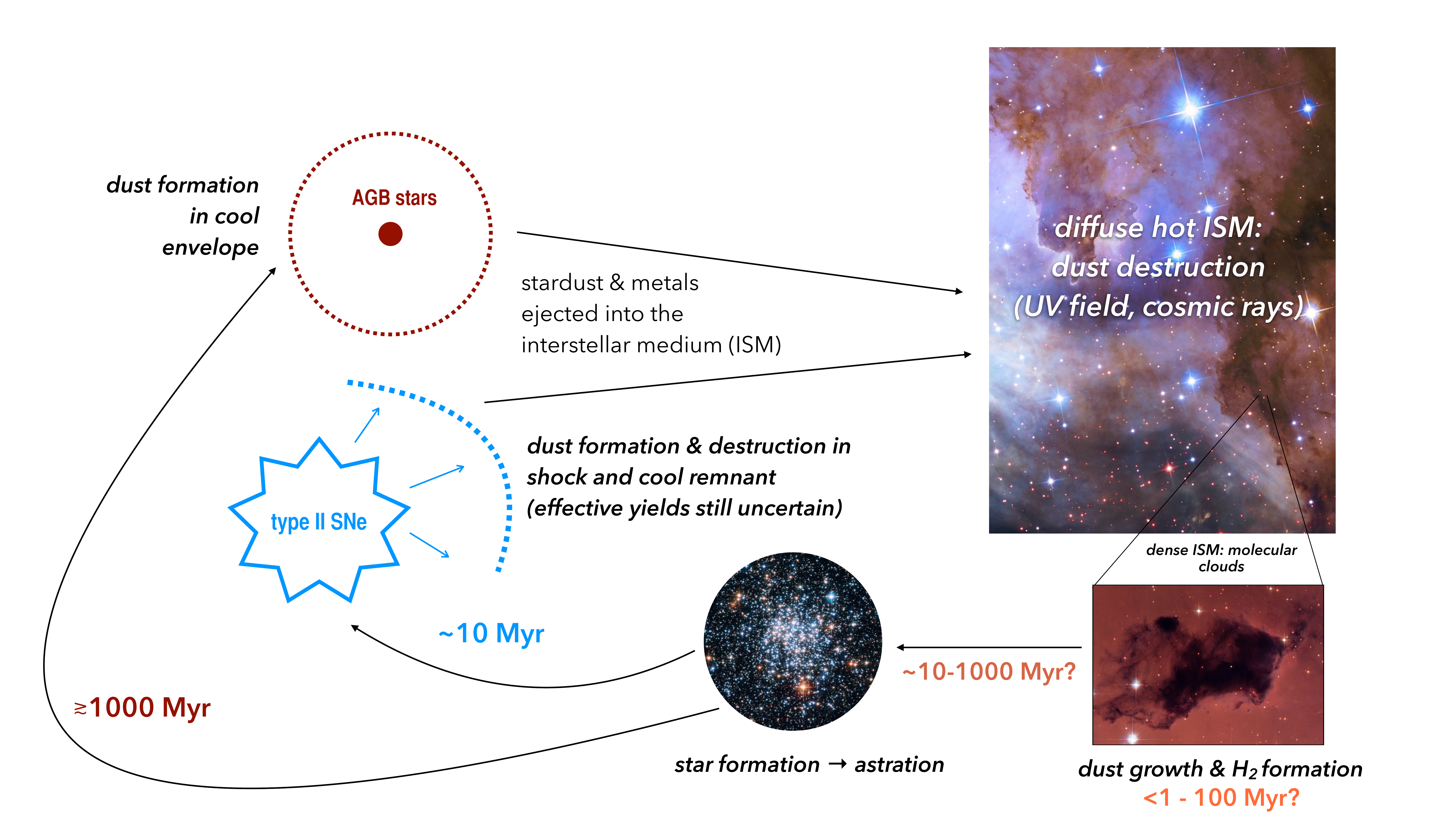}
\caption{A schematic representation of the lifecycle of dust in galaxies (after, e.g., \citealt{Dwek1998,Zhukovska2008,Dwek2011,Asano2013,Zhukovska2014}).}
\label{lifecycle}
\end{center}
\end{figure*}

Having large amounts of dust in early galaxies, when the Universe was less than a billion years old, could be a challenge for our current understanding of how dust forms and evolves. The lifecycle of dust in intimately related with the lifecycle of stars (Fig.~\ref{lifecycle}). In the local Universe, one of the main environments for stardust formation is the cool envelope of AGB stars, however, the timescales for these stars to evolve are of the order of 1 billion years or more (i.e., longer than the Hubble time at $z>6$). Based on stellar evolution timescales, at early ages, the main channel for stardust production has to be supernova dust, however the effective dust yields in supernova are still uncertain even in the local Universe. After being formed in stellar envelopes/supernova remnants, dust grows through coagulation and mantle accretion in dense molecular clouds, and although the timescale and efficiency for this process is still debated, models predict that this ISM growth is only efficient once the gas reaches a `critical metallicity' ($Z_\mathrm{crit}\sim 0.05 - 0.3\,Z_\odot$; e.g., \citealt{Asano2013}, see also \citealt{Galliano2018} for a review). For all these reasons, in galaxies at the epoch of reionization (which presumably have young ages and low metallicities), the main source of dust should be supernova stardust (e.g., \citealt{Gall2011}).

The different channels of dust production and growth also affect the grain size distributions, which in turn affect the dust extinction curves in galaxies (e.g., \citealt{Asano2013b,Aoyama2020}). This could explain the location of high-redshift galaxies in the IRX-$\beta$ diagram. \cite{Mancini2016} post-processed a hydrodynamical galaxy simulation with a detailed chemical and dust evolution model and find that lower stellar mass galaxies at high redshift, which are less chemically evolved, follow a steep SMC-like attenuation curve because their dust is mostly contributed by stellar sources, with less efficient ISM growth than in more metal-rich massive galaxies (see also, \citealt{Graziani2020}). 

Regarding the production of the large dust masses found at high-redshift (REBELS and other ALMA observations find dust masses $M_\mathrm{dust} \gtrsim10^7\,M_\odot$ at $z\simeq7$), models have not reached a consensus yet. Some models find that the dust masses can be explained by high supernova rates and/or yields, or efficient ISM dust growth (e.g., \citealt{Popping2017}), while others find that efficient ISM growth is needed (e.g., \citealt{Mancini2015}). The need for efficient ISM dust growth in early galaxies may be problematic: \cite{Ferrara2016} point out that at high-redshifts, icy dust grain mantles formed in the cold ISM do not survive in the hot diffuse ISM. Recently, \cite{Dayal2022} showed that the DELPHI semi-analytic model can reproduce the dust masses of REBELS sources with minimal ISM dust growth, coupling metal and dust enrichment, and using new stellar yields from \cite{Kobayashi2020}.

This is not a resolved issue. There are still large uncertainties in modelling ISM dust growth and stellar yields. On the observational side, the dust mass measurements of most sources are still uncertain as they rely on limited ALMA continuum measurements. The stellar masses, gas-phase metallicities, and star formation histories of these sources are also critical ingredients in the dust formation and evolution modelling, but so far these measurements are either non-existent (in the case of metallicities) or very uncertain (in the case of the stellar masses and star formation histories; \citealt{Topping2022}). Future observations of the rest-frame optical emission of EoR galaxies with the JWST will be critical to constrain these properties (e.g., \citealt{Schaerer2022,Katz2022}).

\section{A few surprises/challenges from recent ALMA observations}

\subsection{A hidden population of dusty galaxies in the EoR?}

A surprising result to come out of the REBELS survey was the serendipitous detection of two dust-obscured galaxies at $z\simeq7$ that have no optical counterparts in deep HST imaging \citep{Fudamoto2021} -- this implies SEDs and dust attenuations similar to those of the dustiest known galaxies, sub-millimetre galaxies, but in galaxies at much higher redshifts and lower stellar masses than typical SMGs. ALMA observations are not only revealing that there is a variety of dust obscuration at high-redshifts, but that there might be a hidden population of dusty galaxies at the EoR that we did not know about before (similar to optically-dark galaxies seen at lower redshifts; e.g., \citealt{Wang2019,Smail2021}). \cite{Fudamoto2021} estimate that these dusty galaxies could contribute 10 to 25\% of the total cosmic SFR density at $z>6$. This intriguing population needs further study, and these sources are prime targets for JWST follow-up.

\subsection{Spatial offsets between UV/optical and infrared emission}

Some of the REBELS sources show large spatial offsets between the observed ultraviolet and infrared emission -- up to 1.5\,arcsec (or 7.6\,kpc; \citealt{Inami2022}). Similar offsets have also been observed in high-resolution observations of star-forming galaxies at lower redshifts (e.g., \citealt{Hodge2019,Cochrane2021}). These offsets could be a result of the multi-phase ISM -- dust and stars are not homogeneously distributed, with opaque star-forming clumps and a more transparent diffuse component. This could be a challenge for SED fitting methods that assume energy balance (e.g., \citealt{daCunha2008}), as well as for modelling the IRX-$\beta$ relation \citep{Ferrara2022}. More high-resolution follow-up of the dust (with ALMA) and stellar (with JWST) emissions in these sources will be key to better understand these offsets.

\section{Summary \& Open Questions}

Thanks to ALMA we can now detect the dust emission of galaxies all the way into the epoch of reionization. Dust at those early times seems to have different properties than dust in star-forming galaxies at lower redshifts, with some indication of steeper attenuation curves, resembling dust in the SMC rather than in the Milky Way. However, this result depends on assumptions on the dust temperature (used to extrapolate single-band continuum measurements to obtain the total infrared luminosity). Dust in high-redshift galaxies seems to be warmer than dust in lower-redshift galaxies, but not enough to reconcile their infrared excesses with a Milky Way-like attenuation curve, and not as hot as predicted by theoretical models or by extrapolating the temperature-redshift relation calibrated at lower redshift. More direct measurements of the dust emission SEDs in multiple bands for larger samples of galaxies are needed to confirm this result.

The contribution of dust-obscured star formation rate to the cosmic SFR density at $z\gtrsim6$ is non-negligible, and it could be as high as 30\% based on recent REBELS results. This however needs to be confirmed with larger, preferably, blind surveys of dust emission at those redshifts.

ALMA observations show that many $z>6$ UV-selected galaxies have copious amounts of dust. The production of high dust masses in the early Universe still pose a challenge to models. For now there seems to be a consensus in the literature that high supernova yields and/or significant ISM dust growth are needed, but the exact contribution of each process is still debated. Going forward, we need to improve the measurements of the main stellar and dust parameters that need to be compared with models, which at the moment have large uncertainties. JWST will yield more robust stellar masses, star formation histories, and metallicities; ALMA multi-band observations will constrain the dust masses more accurately.

The REBELS survey also finds a new population of dust-obscured galaxies in the epoch of reionization with no optical counterparts -- how significant is this population and what are the properties of these galaxies? Searches for larger samples of these sources with ALMA, and JWST follow-up to observe their stellar emission are needed to address these questions.

Finally, REBELS also provides the first opportunity to compare the UV and IR emission in early galaxies in a systematic way. This shows that spatial offsets between these two emissions are common, which begs the question: how should we deal with decoupled stellar and dust emissions? This may involve revising how we model the SEDs of galaxies. High-resolution ALMA and JWST follow-up of these sources will be crucial to further characterise their ISM and stellar morphologies and study the impact of these offsets.

\acknowledgements
I would like to thank the organizers of this symposium for the invitation to present this talk and my colleagues from the REBELS team for their excellent work in producing the results I reviewed. I also gratefully acknowledge the Australian Research Council as the recipient of a Future Fellowship (project FT150100079) and the ARC Centre of Excellence for All Sky Astrophysics in 3 Dimensions (ASTRO 3D; project CE170100013). I acknowledge that this paper was written on Noongar land, and pay my respects to its traditional custodians and elders, past and present.
\def\aj{AJ}
\def\araa{ARA\&A}
\def\apj{ApJ}
\def\apjl{ApJ}
\def\apjs{ApJS}
\def\apss{Ap\&SS}
\def\aap{A\&A}
\def\aapr{A\&A~Rev.}
\def\aaps{A\&AS}
\def\mnras{MNRAS}
\def\pasp{PASP}
\def\pasj{PASJ}
\def\qjras{QJRAS}
\def\nat{Nature}
\def\physrep{Physics Reports}

\def\aplett{Astrophys.~Lett.}
\def\aas{AAS}
\let\astap=\aap
\let\apjlett=\apjl
\let\apjsupp=\apjs
\let\applopt=\ao


\end{document}